\documentclass[prl,twocolumn,showpacs,amsmath,amssymb,superscriptaddress]{revtex4-2}
\usepackage{dcolumn}
\usepackage{bm,graphicx}
\usepackage{etoolbox}
\usepackage{url}
\usepackage{booktabs}
\usepackage[colorlinks]{hyperref}
\usepackage{breakurl}
\usepackage{color}

\begin{document}

\title{Magnon-polarons in van der Waals antiferromagnet FePS$_3$}

\author{D.~Vaclavkova}
    \thanks{These two authors contributed equally}
    \affiliation{Laboratoire National des Champs Magn\'etiques Intenses, LNCMI-EMFL, CNRS UPR3228, Univ. Grenoble Alpes, Univ. Toulouse, Univ. Toulouse 3, INSA-T, Grenoble and Toulouse, France}

\author{M.~Palit}
    \thanks{These two authors contributed equally}
    \affiliation{School of Physical Sciences, Indian Association for the Cultivation of Science, 2A $\&$ B Raja S. C. Mullick Road, Jadavpur, Kolkata – 700032, India}

\author{J.~Wyzula}
    \affiliation{Laboratoire National des Champs Magn\'etiques Intenses, LNCMI-EMFL, CNRS UPR3228, Univ. Grenoble Alpes, Univ. Toulouse, Univ. Toulouse 3, INSA-T, Grenoble and Toulouse, France}

\author{S.~Ghosh}
    \affiliation{School of Physical Sciences, Indian Association for the Cultivation of Science, 2A $\&$ B Raja S. C. Mullick Road, Jadavpur, Kolkata – 700032, India}

\author{A.~Delhomme}
    \affiliation{Laboratoire National des Champs Magn\'etiques Intenses, LNCMI-EMFL, CNRS UPR3228, Univ. Grenoble Alpes, Univ. Toulouse, Univ. Toulouse 3, INSA-T, Grenoble and Toulouse, France}

\author{S.~Maity}
    \affiliation{School of Physical Sciences, Indian Association for the Cultivation of Science, 2A $\&$ B Raja S. C. Mullick Road, Jadavpur, Kolkata – 700032, India}

\author{P.~Kapuscinski}
    \affiliation{Laboratoire National des Champs Magn\'etiques Intenses, LNCMI-EMFL, CNRS UPR3228, Univ. Grenoble Alpes, Univ. Toulouse, Univ. Toulouse 3, INSA-T, Grenoble and Toulouse, France}
    \affiliation{Department of Experimental Physics, Wroclaw University of Technology, Wybrzeze Wyspianskiego 27, 50-370 Wroclaw, Poland}

\author{A.~Ghosh}
    \affiliation{School of Physical Sciences, Indian Association for the Cultivation of Science, 2A $\&$ B Raja S. C. Mullick Road, Jadavpur, Kolkata – 700032, India}

\author{M.~Veis}
    \affiliation{Laboratoire National des Champs Magn\'etiques Intenses, LNCMI-EMFL, CNRS UPR3228, Univ. Grenoble Alpes, Univ. Toulouse, Univ. Toulouse 3, INSA-T, Grenoble and Toulouse, France}
    \affiliation{Institute of Physics, Charles University, Ke Karlovu 5, Prague, 121 16, Czech Republic}

\author{M.~Grzeszczyk}
    \affiliation{Laboratoire National des Champs Magn\'etiques Intenses, LNCMI-EMFL, CNRS UPR3228, Univ. Grenoble Alpes, Univ. Toulouse, Univ. Toulouse 3, INSA-T, Grenoble and Toulouse, France}
    \affiliation{Institute of Experimental Physics, Faculty of Physics, University of Warsaw, ul. Pasteura 5, 02-093 Warszawa, Poland}

\author{C.~Faugeras}
    \affiliation{Laboratoire National des Champs Magn\'etiques Intenses, LNCMI-EMFL, CNRS UPR3228, Univ. Grenoble Alpes, Univ. Toulouse, Univ. Toulouse 3, INSA-T, Grenoble and Toulouse, France}

\author{M.~Orlita}
    \affiliation{Laboratoire National des Champs Magn\'etiques Intenses, LNCMI-EMFL, CNRS UPR3228, Univ. Grenoble Alpes, Univ. Toulouse, Univ. Toulouse 3, INSA-T, Grenoble and Toulouse, France}
    \affiliation{Institute of Physics, Charles University, Ke Karlovu 5, Prague, 121 16, Czech Republic}


\author{S.~Datta}
    \email{subhanano@gmail.com}
    \affiliation{School of Physical Sciences, Indian Association for the Cultivation of Science, 2A $\&$ B Raja S. C. Mullick Road, Jadavpur, Kolkata – 700032, India}

\author{M.~Potemski}
    \email{marek.potemski@lncmi.cnrs.fr}
    \affiliation{Laboratoire National des Champs Magn\'etiques Intenses, LNCMI-EMFL, CNRS UPR3228, Univ. Grenoble Alpes, Univ. Toulouse, Univ. Toulouse 3, INSA-T, Grenoble and Toulouse, France}
     \affiliation{Institute of Experimental Physics, Faculty of Physics, University of Warsaw, ul. Pasteura 5, 02-093 Warszawa, Poland}

\date{\today}

\begin{abstract}
The hybridization of magnons (spin waves) with phonons, if sufficiently strong and comprising of long wavelength excitations, may offer a new playground when manipulating the magnetically ordered systems with light. Applying a magnetic field to a quasi-2D antiferromagnet, FePS$_3$, we tune the magnon-gap excitation to coincide with the initially lower-in-energy phonon modes. Hybrid magnon-phonon modes, the magnon polarons are unveiled with the demonstration of a pronounced avoided crossing between the otherwise bare magnon and phonon excitations. The magnon polarons in FePS$_3$ are traced with Raman scattering experiments. However, as we show, they also couple directly to terahertz photons, evoking their further explorations in the domain of antiferromagnetic opto-spintronics. The  magnon-phonon coupling is also discussed as a possible reason of the magnon mode splitting observed in the absence of a magnetic field.
\end{abstract}

\maketitle

\section{Introduction}

Research on magnetic solids recently gave rise to a plethora of emerging domains of study, which are motivated by the scientific curiosity to uncover new phenomena and triggered by the possible design of, optional to electronic, spintronic devices \cite{StampsJPD14}. Among current trends in the spintronic developments, there are attempts to exploit antiferromagnetic materials \cite{BaltzRMP18} (instead of ferromagnets) to work with two-dimensional systems \cite{GongScience19,GibertiniNatureNano19}, as well as attempts to manipulate magnetically ordered solids with light \cite{NemecNaturePhys18}. These ideas have stimulated our magneto-optical studies of the representative layered antiferromagnet, FePS$_3$, in which the characteristic magnon-gap excitation appears at a significantly high frequency of a few terahertz \cite{SekinePRB90,WildesJPCM12}. Both the antiferromagnetic order and magnon excitation may survive at the nanoscale, down to the monolayer limit \cite{LeeNL16,ChittariPRB16}.

Magnons (spin waves) and phonons (lattice vibrations) are two relevant, low energy excitations in magnetically ordered systems. The coupling between these modes, central for the present work, has been a subject of numerous theoretical and experimental studies in various ferromagnetic \cite{KamraSSC14,GodejohannPRB20}, antiferromagnetic \cite{SimensenPRB19,LiPRL20}, ferrimagnetic \cite{StreibPRB19,KikkawaPRL16} as well as in multiferroic \cite{PimenovNaturePhys06} materials. Coupling between magnons and phonons affects the dynamical and optical properties of these quasi-particles and appears to be of special importance in emerging areas, such as spin caloritronics \cite{BauerNatureMater12,BozhkoLTP20} or magnon spintronics in conjunction with new developments in terahertz (THz) technology \cite{ChumakNaturePhys15,WalowskiJAP16}.

Coupled magnon-phonon modes, the magnon-polarons, are best evidenced by the observation of the repulsion (avoided crossing) of the otherwise bare phonon and magnon excitations when they are brought to coincide. Such concurrency is customarily expected when the (strongly) linearly dispersing acoustic phonons cross the quadratically dispersing magnons and/or when the dispersing magnons intersect the weakly dispersing (optical-like) phonons, notably, at certain nonzero k-wavevectors in both cases. Magnon polarons can then be evidenced with neutron scattering experiments and/or techniques of surface acoustic waves, both providing an access to $k \neq 0$ excitations \cite{GodejohannPRB20,ManPRB17,SukhanovPRB19,DreherPRB12}. Magnon and phonon excitations can also be probed by means of optical techniques which are easily operational in conjunction with the application of high magnetic fields. Such methods provide an access to yet another regime, to substantiate evidence of a possible coupling between the $k=0$ magnon and phonon modes with dispersions that do not cross at any wavevector.

\begin{figure*}[t]
      \includegraphics[width=1\textwidth]{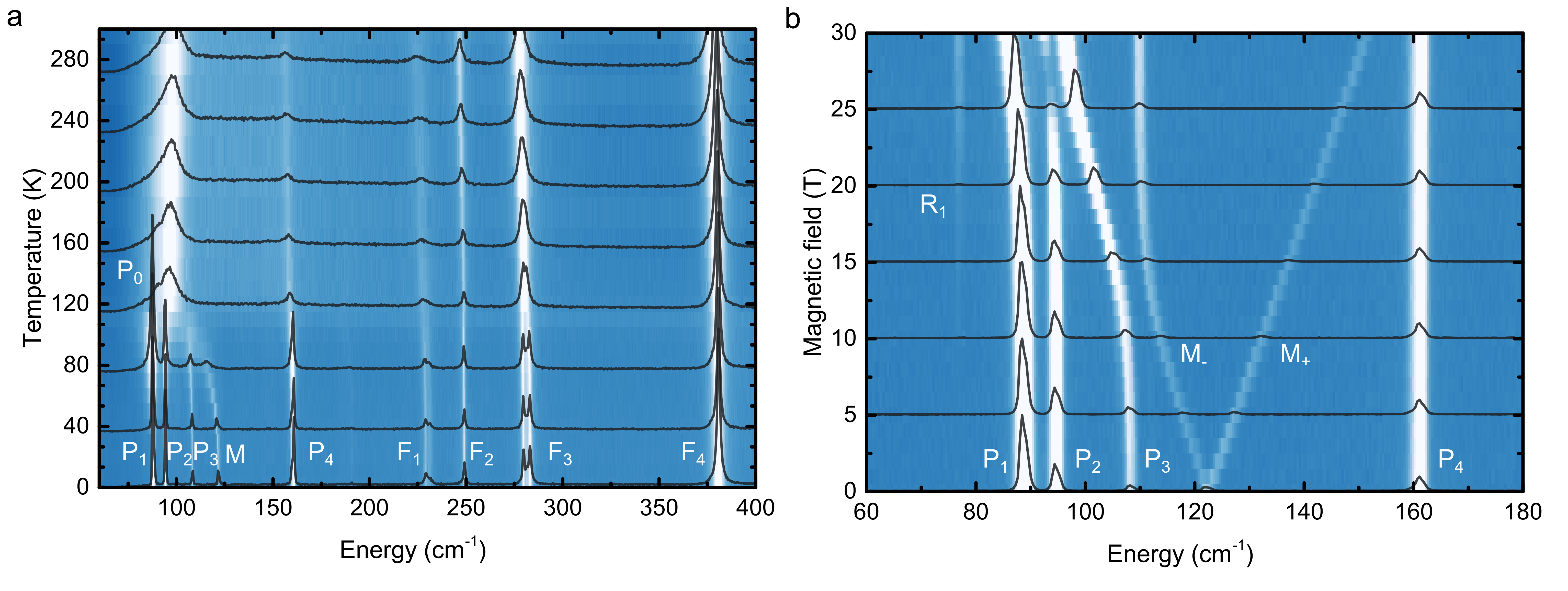}
      \caption{\label{Fig1} (a) False-color map of the Raman scattering response of FePS$_3$ antiferromagnet, together with a few selected characteristic spectra, measured as a function of temperature in the range from 4.2~K up to 300~K. $P_i$ ($i=1\ldots4$) and $F_i$ ($i=1\ldots4$) resonances are due to phonon modes and the $M$-feature corresponds to the magnon-gap excitation. (b) False-color map of the evolution of the low-temperature ($T = 4.2$~K) magneto-Raman scattering response of FePS$_3$ with an applied magnetic field oriented perpendicular to the plane of the layers, together with a few selected spectra, in the spectral region from 60 to 180 cm$^{-1}$. The coupling between the lower magnon branch $M_-$ and three $P_i$ ($i=1\ldots3$) phonon modes is clearly observed. An additional $R_1$ phonon mode is activated at high magnetic fields. The energy of the upper $M_+$ magnon branch smoothly develops with the magnetic field and does not reach the energy of the $P_4$ phonon in the range of magnetic fields investigated.}
\end{figure*}

Here we unveil the magnon polarons in the FePS$_3$ antiferromagnet with optical spectroscopy experiments, magneto-Raman scattering and far-infrared/terahertz (FIR/THz) magneto-transmission measurements, thus naturally probing $k = 0$ excitations. By applying a magnetic field, we split the magnon-gap excitation into two components and drive the lower energy one to intersect the three characteristic Raman- and FIR-active phonon modes, but instead, we observe a series of pronounced anticrossing events. The analysis of the observed anticrossing pattern allows us to estimate the strength of the magnon-phonon coupling in the limit of the magnetic material in which the magnon mode is coupled to lower in energy, $k = 0$ phonons. We propose that this coupling alters the optical selection rules for magnon and phonon excitations what, in particular, implies their effective activation with THz photons. The present work complements the previous~\cite{McCrearyPRB20} and very recent~\cite{LiuPRL21} magneto-Raman scattering studies of FePS$_3$ antiferromagnet with, in particular, the far-infrared magneto-spectroscopy data and observation of the zero field splitting of the magnon gap, the latter likely also induced by the magnon-phonon coupling.

Although FePS$_3$ is among the best-known layered antiferromagnets within the large family of transition metal phosphorus trichalcogenides (TMPT) \cite{GrassoLRdNC02}, the exact rules governing the spin ordering in this material continue being revisited~\cite{WildesJPCM12,TaylorJSSC73,KurosawaJPSJ83,JoyPRB92,RulePRB07,LanconPRB16,WildesJAP20}. This includes very recent reports invoking possible effects of spin-lattice coupling \cite{WildesPRB20,WildesPRB20}. Below the N\'eel temperature of $T_N = 120$~K, FePS$_3$ is generally considered as a good example of a two-dimensional antiferromagnet~\cite{WildesJPCM12,TaylorJSSC73,KurosawaJPSJ83,JoyPRB92,RulePRB07,LanconPRB16,WildesJAP20}, even in its bulk form which is composed of weakly bound, via van der Waals forces, layers with Fe$^{2+}$ (S = 2) spins arranged on a honeycomb lattice~\cite{GrassoLRdNC02,OuvrardMRB85}. The interlayer spin-spin exchange terms are weak and the antiferromagnetic order, with Fe$^{2+}$ spins aligned along the direction perpendicular to the layers plane, is largely governed by intralayer exchange integrals and the strong term of the single Fe$^{2+}$ ion anisotropy~\cite{LanconPRB16}. This latter term justifies the Ising-type notion for antiferromagnetism in FePS$_3$~\cite{WildesJPCM12,TaylorJSSC73,KurosawaJPSJ83,JoyPRB92,RulePRB07,LanconPRB16,WildesJAP20}. It is also the reason for the relatively large energy of the magnon gap, the zero wavevector ($k=0$) excitation of the lower energy branch of the spin-waves, in this material. The spin-wave/magnon dispersion relations in FePS$_3$ have been widely studied with neutron scattering ~\cite{WildesJPCM12,RulePRB07,LanconPRB16,WildesJAP20}. The magnon gap in this antiferromagnet at low temperature has been identified in Raman scattering experiments with a characteristic signature at $E_M \approx 122$~cm$^{-1}$~\cite{SekinePRB90,McCrearyPRB20}.

\begin{figure*}[t]
      \includegraphics[width=0.75\textwidth]{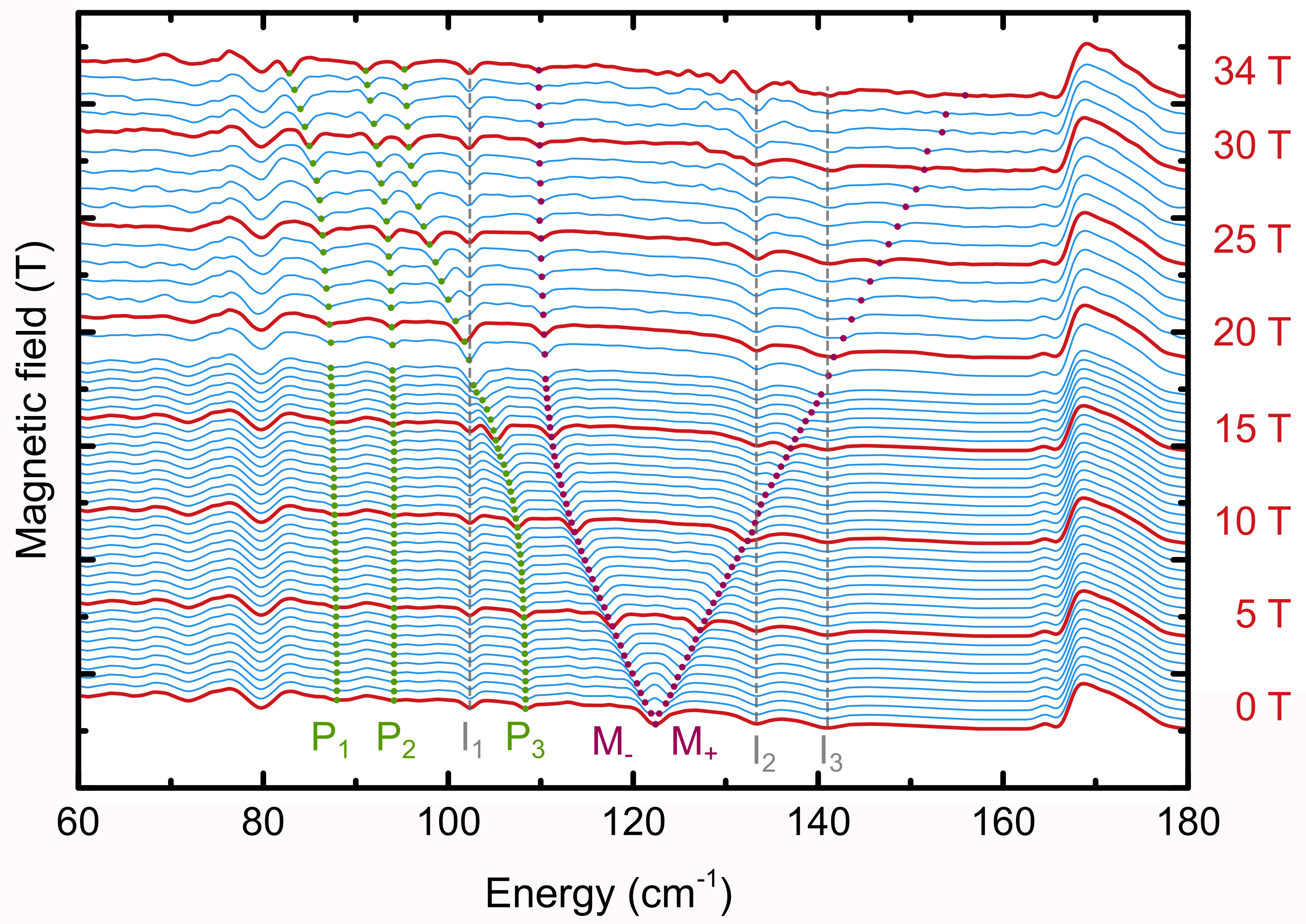}
      \caption{\label{Fig3} Low temperature ($T = 4.2$~K) far-infrared transmission spectra for selected values of the magnetic field applied perpendicular to the layer planes, in the spectral region 60 - 180 cm$^{-1}$. Transmission minima denoted as $P_i$ ($i=1\ldots3$) and $M_+$/$M_-$ have their counterparts in resonances observed in Raman scattering spectra. $I_i$ ($i=1\ldots3$) transitions are only visible in FIR transmission spectra. The pronounced minimum at $\sim$ 80 cm$^{-1}$ is due to absorption resonance in the polyethylene foil used in our experimental set-up to filter the high frequency radiation.}
\end{figure*}

\section{Experimental Details}

The investigated samples consisted of relatively thick flakes (thickness of $\sim$1-10~$\mu$m) extracted from bulk FePS$_3$ crystals and deposited on Si/SiO$_2$ or Si substrates. The surfaces of the flakes have been ``cleaned'' by the ``exfoliation'' method before each experimental run to obtain a surface with good optical quality. Either commercially available (HQ Graphene) or home-grown FePS$_3$ crystals were utilized. The latter crystals were grown by the chemical vapor transport method in two zone furnaces following an established method~\cite{DuACSNano16}. All samples have been initially tested with room temperature Raman scattering measurements, and they all showed practically the same characteristics.

The temperature dependent Raman scattering response was measured at zero magnetic field in a continuous flow cryostat mounted on $x$-$y$ motorized positioners. The sample was placed on the cold finger of the cryostat and excited with the 515~nm line of a continuous-wave laser diode. The excitation light was focused by means of a 50$\times$ long-working distance objective with a 0.5 numerical aperture producing a spot of about 1~$\mu$m and the scattering signal was collected via the same objective. Low-temperature magneto-Raman experiments were performed in the back scattering configuration with the magnetic field applied perpendicularly to the $ab$-plane of our sample. We used the Faraday geometry, i.e., the magnetic field is parallel to the light propagation direction. Measurements were carried out with magnetic fields up to 30~T using a free-beam-optics arrangement. The sample was placed on top of a $x$-$y$-$z$ piezo-stage (kept in gaseous helium at $T = 4.2$~K), inserted into a magnet and was excited using a 515.1~nm line of a continuous wave laser diode (2.41~eV photon energy). The emitted light was dispersed with a 0.7~m long monochromator and detected with a CCD camera.

Far-infrared transmission experiments were carried out on a macroscopic FePS$_3$ bulk sample (surface size ~10~mm$^2$) which was kept in the He exchange gas at the temperature of $T=$4.2~K and placed in a superconducting solenoid (magnetic field up to 18~T) or a resistive coil (magnetic field up to 34~T). The magnetic field was oriented perpendicular to the $ab$-plane of the FePS$_3$ crystal. To measure infrared magneto-transmission, the radiation from a globar source was modulated by a Bruker Vertex 80v Fourier-transform spectrometer, delivered to the sample via light-pipe optics and detected by a composite bolometer placed directly behind the sample. All measurements were performed in the Faraday geometry. The presented transmission spectra were normalized by the transmission of a 1~mm pinhole measured at each magnetic field, thus correcting for magnetic-field-induced variations in the response of the bolometer.

\section{Experimental Results}

With the results presented in Fig.~\ref{Fig1}a, we recall the temperature evolution of the Raman scattering response typically observed in FePS$_3$ crystals~\cite{SekinePRB90,McCrearyPRB20,ScagliottiPRB87,LeeNL16,GhoshPRB21}. The $F_i$ ($i = 1, 2, 3, 4$) features observed above 200 cm$^{-1}$ are largely due to molecular-like vibration of the (P$_2$S$_6$)$^{4-}$ anion unit (which together with two Fe$^{2+}$ cations form the simplified unit cell of FePS$_3$) and are pretty common for all TMPT compounds~\cite{BalkanskiJPCSSP87,HashemiJPCC17}. Instead, the Raman scattering peaks seen at lower energies ($P_i$ features) are expected to be due to phonons which include the vibration of Fe$^{2+}$ ions~\cite{LeeNL16,BalkanskiJPCSSP87} while the M feature is now well recognized~\cite{McCrearyPRB20} as due to the magnon-gap excitation. As previously reported~\cite{SekinePRB90,McCrearyPRB20,LeeNL16} and illustrated in Fig.~\ref{Fig1}a, the magnon peak M as well as all $P_i$ peaks are sensitive to the magnetic ordering: when temperature is raised above the N\'eel temperature of $T_N\approx 120$~K, the intensity of the $P_4$ resonance drops down abruptly, whereas $P_i$ ($i=$1,2,3) and $M$ peaks merge together into a broad $P_0$ feature. The $P_4$ peak is commonly associated to a $A_g$/$B_g$ phonon from the center of the Brillouin zone of the FePS$_3$ crystal~\cite{McCrearyPRB20,LeeNL16,GhoshPRB21,BalkanskiJPCSSP87,Wang2DM16,KargarACSNano20}. Instead, the identification of phonons associated with $P_i$ ($i=$0,1,2,3) resonances is less conclusive and we will comment more on this issue later on.

In the following we focus our attention on the low energy spectral range (60-180 cm$^{-1}$) and low temperature regime (4.2~K) and examine the Raman scattering and far-infrared transmission spectra of our FePS$_3$ crystal, measured as a function of the magnetic field applied perpendicularly to the layer planes, i.e., along the direction of Fe$^{2+}$ ions' spin alignment. The results of magneto-Raman scattering measurements are illustrated in Fig.~\ref{Fig1}b. In accordance to the previous study performed at low magnetic fields~\cite{McCrearyPRB20}, the very first effect of the application of a magnetic field is the splitting of the magnon peak into two, $M_+$ and $M_-$ components. This energy splitting, approximately linear with the magnetic field (B) in the range of low fields, scales as $2g\mu_B B$ where $\mu_B$ stands for the Bohr magneton and, to the first approximation, we find $g\approx2$ for the effective $g$-factor, in line with the previous report~\cite{McCrearyPRB20}.

The present work highlights the effects observed at high magnetic fields, when the $M_-$ magnon branch is tuned in the spectral range of three low energy $P_i=$1,2,3-phonons. As can be seen in Fig.~\ref{Fig1}b, the $M_-$ magnon excitation does not intersect any of the $P_1$, $P_2$ and $P_3$ phonons and instead a characteristic pattern of avoided crossing events is observed, in line with recent high-field magneto-Raman study~\cite{LiuPRL21}. A simple inspection of the raw data leads us to conclude that the $M_-$ magnon effectively couples to all three $P_1$, $P_2$ and $P_3$ phonons. Besides that, we observe at high magnetic fields (above 14~T) the activation of an additional Raman scattering peak, presumably due to another phonon excitation. The energy position of this additional excitation (marked as $R_1$ in Fig.~\ref{Fig1}b) does not, however, change with the magnetic field what prevents us to firmly conclude about its potential strong coupling to the magnon mode. As for the upper, $M_+$ component of the magnon mode, we note its smooth blue shift, approximately linear with the magnetic field. There are no Raman active phonon modes in the spectral range covered by the $M_+$ magnon component which only approaches the $P_4$ phonon at the highest available magnetic fields, but still not sufficiently close to let us conclude about their possible hybridization. It is still worth noting that the extrapolated crossing of the $M_+$ - magnon branch and the $P_4$ phonon is expected at $B\approx40$~T, that is at the field strength at which the phase transition of the FePS$_3$ ground state has been recently anticipated from magnetization measurements~\cite{WildesPRB20}.

The results presented in Fig.~\ref{Fig3} demonstrate that several excitations among those traced with Raman scattering do also directly couple to light, giving rise to absorption resonances observed in FIR magneto-transmission measurements, which is an experimental technique different from Raman scattering used in previous magneto-optical studies~\cite{McCrearyPRB20,LiuPRL21}. Tracking these resonances (minima/dips in transmission spectra) as a function of the magnetic field we are able to reproduce the characteristic pattern of avoided crossings of the $M_-$ magnon branch with the $P_1$, $P_2$ and $P_3$ phonons. In contrast to $M_-$/$M_+$ and to the $P_3$ excitations that can be clearly observed at anyu value of the magnetic field, the transmission minima related to $P_1$ and $P_2$ phonons are rather weakly pronounced at low magnetic fields. They are barely visible in the raw data, but still present even in the spectrum measured in the absence of a magnetic field as deduced when examining the measured magneto-transmission response in more details (see Ref.~\cite{SM}). On the other hand the transmission minima, marked as $I_1$, $I_2$ and $I_3$ in Fig.~\ref{Fig3} are assigned to other absorption resonances in the FePS$_3$ crystal. These resonances are not Raman active and do not couple to magnon modes (cross either $M_-$ or $M_+$ magnon branch) and we also note that the $R_1$ and $P_4$ resonances, otherwise observed in Raman spectra are not optically active, not visible in magneto-transmission spectra. For the sake of completeness, let us note that another, relatively weak, spectral feature with the antiferromagnetic-resonance-like splitting appears in the FIR magneto-transmission response at higher energies (around 318~cm$^{-1}$, see Fig.~S3 in~Ref.~\cite{SM}). We assign this mode, in agreement with neutron scattering experiments~\cite{LanconPRB16}, to the $k=0$ excitation of the upper magnon mode, expected in antiferromagnets with four moments in the magnetic unit cell. Nevertheless, as we have checked, this upper magnon mode is hardly apparent in the Raman scattering response.

\begin{figure}[t]
      \includegraphics[width=0.45\textwidth]{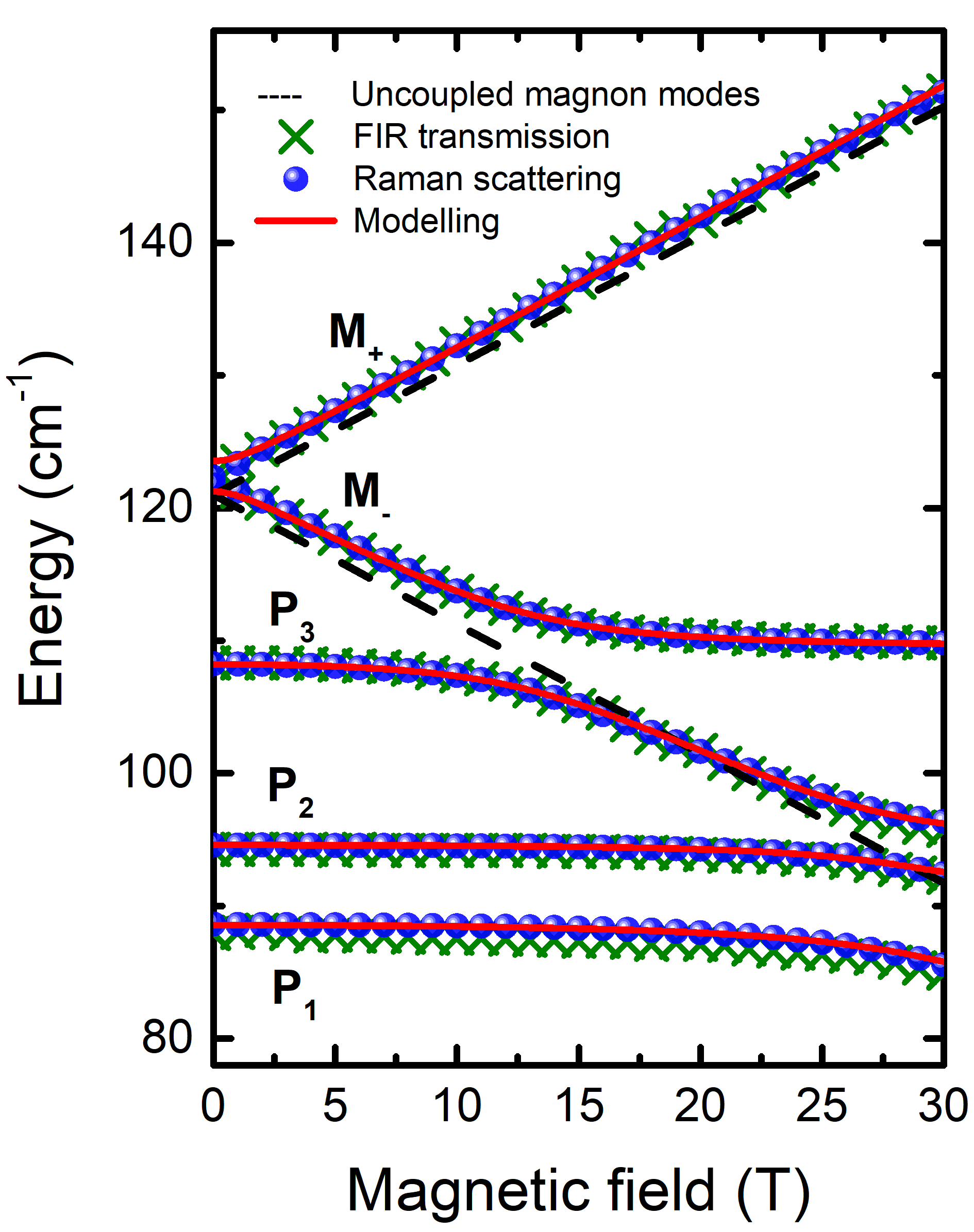}
      \caption{\label{Fig4} Magnetic field dependence of the energy positions of hybrid magnon-phonon modes. Full circles represent the experimental data extracted from magneto-Raman scattering measurements. This data is reproduced with solid lines, following the theoretical modelling described in the text. Crosses account for the energy positions of transitions observed in FIR transmission spectra. The resonances observed in Raman scattering and FIR transmission spectra overlap within the experimental error. Dashed lines show the field dependence of energy positions of magnon modes without coupling to phonons.}
\end{figure}

The $P_1$, $P_2$, and $P_3$ phonon modes which couple to the magnon excitation are not easily identifiable and this also applies to other ($R_1$, $I_1$, $I_2$, $I_3$) resonances observed in our spectra. All these resonances are traced with optical experiments and are thus naturally associated to $\Gamma$-point excitations of FePS$_3$ in its antiferromagnetic phase. However, even for those basic phonons, no consensus exists between different reports of the calculated phonon dispersions in FePS$_3$~\cite{LeeNL16,HashemiJPCC17,BernasconiPRB88,Wang2DM16,KargarACSNano20}. All these reports predict the presence of only very few $\Gamma$-phonons in the low energy range (60-180~cm$^{-1}$), when considering the crystal in the paramagnetic phase. In addition to the $P_4$ Raman peak, only the $P_0$ feature observed at high temperatures (see Fig.~\ref{Fig1}a) are commonly associated with the calculated phonon of the $A_g$/$B_g$ or $E_g$ symmetry~\cite{LeeNL16,HashemiJPCC17,BernasconiPRB88,Wang2DM16,KargarACSNano20}. A specific broadening effect and/or an unresolved multicomponent character of the $P_0$ feature remains to be clarified. The unit cell of the FePS$_3$ crystal is, however, enlarged in the magnetically ordered phase, in the direction across as well as along the planes~\cite{SekinePRB90,LeeNL16}. The appearance of multitude of optically active low energy phonons in the antiferromagnetic phase of FePS$_3$ is expected to be a consequence of the zone folding, in particular of the $M$-point onto the $\Gamma$-point. Apparent phonon energies maybe also affected by an additional deformation of the unit cell of FePS$_3$ at low temperatures~\cite{MurayamaJAP16}. Regrettably, the available results~\cite{LeeNL16,HashemiJPCC17} of calculated phonon modes in the antiferromagnetic phase of FePS$_3$ do not permit a definite identification of the observed resonances. In any case, the $P_i=$1,2,3 phonon modes being central for the present work must be associated with the in-plane motion of Fe$^{2+}$ ions, as they effectively couple to the innately in-plane spin-waves in FePS$_3$. Several Raman-active phonon modes with such symmetry have been predicted to appear in the energy range 70-100~cm$^{-1}$, in calculations limited to a single FePS$_3$ layer~\cite{LeeNL16}. These optical-like and/or ``folded'' phonons are not expected to cross the magnon excitation at any $k$-vector. Their hybridization with the magnon mode is hardly visualized with, for example, conventional neutron scattering experiments, but possible in our studies profiting of the application of magnetic fields~\cite{HayJAP69}. Anticipating new theoretical approaches to calculate the phonon spectra of the FePS$_3$ antiferromagnet, we list, in Table~\ref{tab:my_label}, the characteristic energies of resonances as they appear in the low energy range of the measured Raman and far-infrared absorption spectra. Attempts have been also undertaken to measure the magneto-Raman scattering spectra of FePS$_3$ in the configuration of the magnetic field applied along the crystal planes (Voigt geometry). The results of such experiments are inspiring (see Fig.~S4 in~Ref.~\cite{SM}), although hardly conclusive since surfaces of our crystals are not perfectly flat what prevents the arrangement of a well-defined Voigt geometry (remaining out-of-plane component of the magnetic field).

\section{Theoretical modelling of hybrid magnon-phonon modes}

The effects of magnon-phonon coupling, clearly apparent even in our raw magneto-spectroscopy data (see Fig.~\ref{Fig1}b), are now examined in more details. The central positions of the Raman scattering peaks and the infrared transmission dips associated with the $P_1$, $P_2$, $P_3$ and $M_+$/$M_-$ resonances are plotted as a function of the magnetic field, in Fig.~\ref{Fig4}. To reproduce these data, we refer to a generic theoretical approach to the problem of magnon-phonon coupling~\cite{SimensenPRB19,WhitePR65} which, in its complete form, accounts for possible interactions between all relevant (dispersing with wavevector $k$) branches of magnon and phonon modes present in a magnetically ordered system overall described within the Heisenberg formalism~\cite{LanconPRB16}. This general approach is here simplified down to a phenomenological model. We neglect modes' dispersions and impose a coupling only between $k=0$ excitations. Furthermore, only the interactions which clearly appear in our data are considered, i.e., the interaction between ($M_+$/$M_-$) magnon modes and three $P_1$, $P_2$ and $P_3$ phonon excitations. All other phonon modes, as well as the upper magnon mode at significantly higher energy ($\sim$ $318$ cm$^{-1}$) have been neglected. In its matrix form, the resulting magnon-phonon interaction Hamiltonian is then given by:

\begin{equation}
\label{Ham}
    H = X^\dag H' X
\end{equation}
where
\begin{equation}
\label{Vector}
    X = [ \alpha_{M_+}, \alpha_{M_-}, \zeta_{P_1}, \zeta_{P_2}, \zeta_{P_3}, \alpha^\dag_{-M_+}, \alpha^\dag_{-M_-}, \zeta^\dag_{-P_1}, \zeta^\dag_{-P_2}, \zeta^\dag_{-P_3}]
\end{equation}
is a vector composed of annihilation and creation bosonic operators
$\alpha_{M_+}$, $\alpha_{M_-}$, $\zeta_{P_1}$, $\zeta_{P_2}$, $\zeta_{P_3}$
associated, correspondingly, with the magnetically split magnon gap excitation $\alpha_{M_+}$, $\alpha_{M_-}$ and three (dispersionless) $P_1$, $P_2$ and $P_3$ phonons.
\begin{equation}
\label{HMatrix}
    H' = \begin{bmatrix}
F & G\\
G & F,
\end{bmatrix}
\end{equation}
where
\begin{equation}
\label{FMatrix}
    F = \begin{bmatrix}
E_{M_+}^0 & 0 & \lambda_1 & \lambda_2 & \lambda_3\\
0 & E_{M_-}^0 & \lambda_1 & \lambda_2 & \lambda_3\\
\lambda_1  & \lambda_1  & E_{P_1}^0 & 0 & 0\\
\lambda_2  & \lambda_2  & 0 & E_{P_2}^0 & 0\\
\lambda_3  & \lambda_3  & 0 & 0 & E_{P_3}^0\\
\end{bmatrix}
\end{equation}
and
\begin{equation*}
\label{FMatrix}
    G = \begin{bmatrix}
0 & 0 & \lambda_1 & \lambda_2 & \lambda_3\\
0 & 0 & \lambda_1 & \lambda_2 & \lambda_3\\
\lambda_1  & \lambda_1  & 0 & 0 & 0\\
\lambda_2  & \lambda_2  & 0 & 0 & 0\\
\lambda_3  & \lambda_3  & 0 & 0 & 0\\
\end{bmatrix}
\end{equation*}

 $E^0_{M+/-}$ are the energies of the bare (uncoupled) $M_{_+/_-}$ magnon modes, which are supposed to follow a linear, with the magnetic field B, dependence: $E^0_{M_+/_-} = E^0_M \pm g\mu_B B$ where $E^0_M$
denotes the bare magnon gap at $B = 0$. Similarly, $E^0_{P_1}$, $E^0_{P_2}$ and $E^0_{P_3}$ denote the bare energies of $P_i$ ($i =$~1,2,3) phonons. Three parameters, $\lambda_1$, $\lambda_2$ and $\lambda_3$, account for possible different strengths of magnon-phonon coupling to each $P_1$, $P_2$ and $P_3$ phonon excitations. Each of these phonons is however coupled with the same strength to both $M_-$ and $M_+$ magnon branches and each $\lambda_i$ parameter is assumed to be independent of the strength of the applied magnetic field.

The Hamiltonian \ref{Ham} was diagonalized using the Bogoluybov transformation~\cite{SimensenPRB19,WhitePR65} in order to obtain the field dependent energies of the coupled magnon-phonon modes. Those were compared with the experimental data and the parameters $E_M^0$, $E_{P_1}^0$, $E_{P_2}^0$, $E_{P_3}^0$, $\lambda_1$, $\lambda_2$, $\lambda_3$ and $g$ adjusted for the best agreement using the least square method. The resulting simulation together with the experimental data is shown in Fig.~\ref{Fig4} and the values of fitting parameters are listed in Table~\ref{tab:my_label}.

\begin{table*}[t]
    \centering
    \begin{tabular}{cccccccccccccc}
         \bottomrule
         & \multicolumn{4}{c}{\textbf{Coupled modes}} &\hspace{5mm} & \multicolumn{5}{c}{\textbf{Other resonances}}\\\cmidrule{2-5}\cmidrule{7-11}

         & M$_{\mathrm{agnon}}$ & P$_1$ & P$_2$ & P$_3$ & & R$_1$ & I$_1$ & I$_2$ & I$_3$ & P$_4$ \\
         \cmidrule{2-5}\cmidrule{7-11}
         \multicolumn{11}{c}{}\\

         \multicolumn{4}{c}{\textbf{Experiment (B=0T)}}& & \multicolumn{5}{c}{}\\
         \cmidrule{1-5}\cmidrule{7-11}
         Energy (cm$^{-1}$)& 122.0 &  88.6 & 94.6 & 108.5 & & 77.1 & 102.3 & 133.3 & 141.0 & 161.0 \\
         \cmidrule{1-5}\cmidrule{7-11}
         Activity&  R/IR &  R/IR & R/IR & R/IR & & R & IR & IR & IR & R \\
         \cmidrule{1-5}\cmidrule{7-11}
         \multicolumn{11}{c}{}\\

         \multicolumn{4}{c}{\textbf{Modelling for g-factor =2.15}}& & \multicolumn{5}{c}{}\\\cmidrule{1-5}

         Energy (cm$^{-1}$)& 121.2 &  89.6 & 95.0 & 109.8 & & \multicolumn{5}{c}{} \\\cmidrule{1-5}

         Coupling (cm$^{-1}$)&   &  3.6 & 2.4 & 3.1 & & \multicolumn{5}{c}{} \\
         \bottomrule

    \end{tabular}
    \caption{(upper part). Characteristic energies of resonances observed in Raman scattering (R) and FIR transmission (IR) spectra measured in the absence of magnetic field. (bottom part) Set of parameters (``bare'' energies, coupling constants, and ``bare''  $g$-factor) used to reproduce the magnetic field evolution of hybrid magnon-phonon modes (see Fig.~\ref{Fig4}).}
    \label{tab:my_label}
\end{table*}

\section{Discussion}

As shown in Fig.~\ref{Fig4}, our simplified modelling reproduces the observed energy pattern of the avoided crossings of the $M_-$ magnon branch and three $P_1$, $P_2$ and $P_3$ phonons. The deduced coupling strengths ($\lambda_i$ - lambda parameters) are similar for the hybridization of the magnon with both $P_1$ and $P_3$ phonons and somewhat weaker in case of the $P_2$ phonon (see Table~\ref{tab:my_label}). Both ($M_-$,$P_3$) and ($M_-$,$P_2$) pairs are brought into a strong coupling regime when the $M_-$ magnon and the respective phonons tend to coincide at certain values of the magnetic field: the separation between hybrid modes always surpasses their spectral widths (see Fig.~\ref{Fig1}b). The same can be expected for the pair of ($M_-$, $P_1$) modes, viewing the extracted value of the $\lambda_1$ parameter.

\begin{figure*}
      \includegraphics[width=0.65\textwidth]{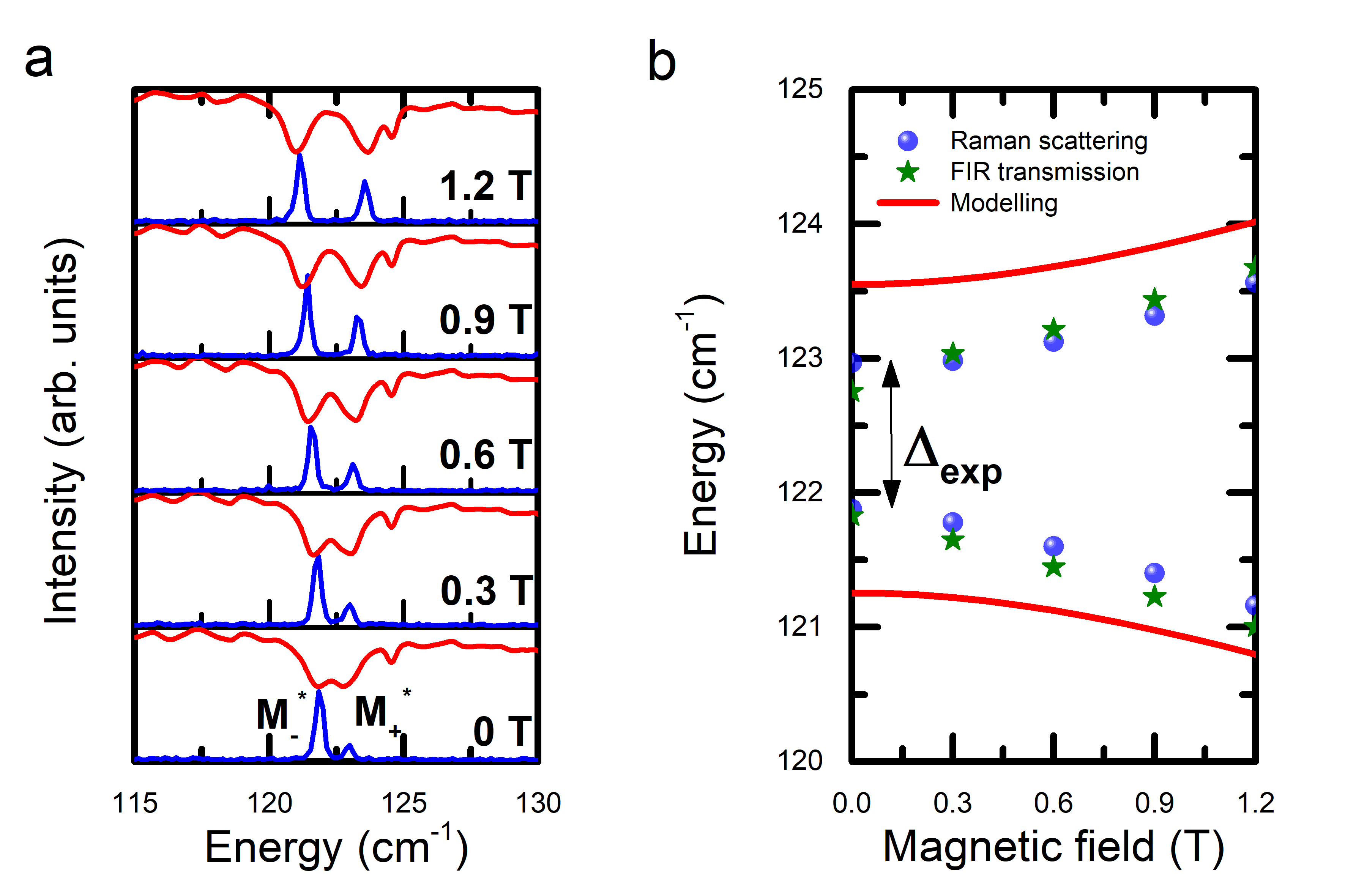}
      \caption{\label{Fig5} a) High resolution magneto-Raman scattering and FIR transmission spectra of magnon resonance measured at low magnetic fields from 0 to 1.2~T. The transmission minimum, slightly above $M_{+}^*$ resonance is only visible in FIR spectra measured under high resolution and is unaffected by the magnon excitation. b) Energies of the two components of the magnon mode as a function of the magnetic field, extracted from the spectra shown in (b) together with the results of data simulation (solid lines) applied to reproduce the evolution of hybrid magnon-phonon modes in the full range of magnetic field investigated (see Fig.~\ref{Fig4}).}
\end{figure*}

The coupling between $M_+$/$M_-$ and $P_i$ ($i=$1,2,3) modes also persists in the absence of magnetic field. As for the phonon modes, this coupling is reflected by the red shift of the experimentally observed resonances with respect to the undressed modes (see Table~\ref{tab:my_label}). The ``renormalization'' of the phonon modes is rather small, at the level of few percents of their apparent energies, but still indicative of the mode hybridization, even in the absence of magnetic field. The magnon-phonon coupling affects also the magnon mode, but in a different way than phonons, i.e., by lifting the two-fold degeneracy of this mode at zero magnetic field (see Fig.~\ref{Fig5}). The zero-field splitting $\Delta$ of the magnon mode, into $M_{+}{^*}$ and $M_{-}{^*}$ components, is not resolved in the experimental data presented in Figs.~\ref{Fig1}b and~\ref{Fig3}. But it is clearly apparent in the spectra shown in Fig.~\ref{Fig5}a, and measured with higher spectral resolution in the range of small magnetic fields. In the frame of our model the lifting of the double degeneracy of the magnon mode is a consequence of its hybridization with phonons of lower energies: two initially degenerate magnon states split into their symmetric and antisymmetric combinations. The energy of lower-energy antisymmetric component $M_{-}{^*}$ is in fact not affected by the coupling and corresponds to the bare magnon mode. The upper symmetric $M_{+}{^*}$ component is shifted up in energy by an amount of $\Delta$. As illustrated in Fig.~\ref{Fig5}b, $\Delta \approx 1$~cm$^{-1}$ is derived from the experiment. This value is smaller, by a factor of two, from the one extracted from our calculation what can signify the limitation of our simple phenomenological model. We have, for example, checked that imposing certain coupling (strength up to $\lambda = 2$~cm$^{-1}$) of the magnon mode with the higher energy $P_4$ phonon does not practically change the quality of data simulation presented in Fig.~\ref{Fig4}. But it can suppress the amplitude of $\Delta$ closer to its experimental value. It is logical to assume that the zero-field spitting we observe for the magnon-gap excitation in FePS$_3$ is a consequence of the apparent magnon-phonon interaction in this material. We note, that such splitting may also be driven by other mechanisms, including those referring to purely magnetic interactions~\cite{KobetsLTP09,ShenPRL20}. This, in combination with our findings calls for further revision of antiferromagnetism in FePS$_3$, with theoretical studies in particular.

Having established the characteristic behavior of the energy spectrum of coupled magnon-phonon modes, we now turn to the discussion of the uncommon optical selection rules. They mediate the observation of these modes in our spectra. Focusing first on the magnon excitation and on its Raman scattering response, we confirm that the selection rules, established long time ago~\cite{FleuryPR69} for the conventional MnF$_2$ and FeF$_2$ antiferromagnets, cannot strictly be applied to the case of FePS$_3$~\cite{McCrearyPRB20}. Certain dichroism in the magneto-Raman scattering remains apparent. As demonstrated in Ref.~\cite{SM}, the observation of the $M_+$ ($M_-$) branch is favored in the configuration of circularly cross-polarized (co-polarized) beams of the excitation and scattered light. On the other hand, as also shown in~Ref.~\cite{SM}, the $M_{-}{^*}$/$M_{+}{^*}$ zero-field components of the magnon excitation display the characteristic Raman selection rules when probed under conditions of differently oriented linear polarization of the excitation and scattered light. Other observations which bring our particular attention are: (a) magnon and $P_i=$1,2,3 phonon resonances are apparent in both Raman scattering as well as in photon absorption processes. Whereas the conventional selection rules are usually different for those two processes, at least in reference to phonon resonances, (b) magnon excitation raises a strong absorption resonance (as seen in FIR transmission spectra), although the unit cell of FePS$_3$ is commonly assumed to preserve the inversion symmetry in the antiferromagnetic phase~\cite{KrivoruchkoLTP12}. Thus the magnon-absorption process in this material might not be expected to be active within the electric-dipole approximation. (c) in the absence of magnetic field, i.e., in the regime of weak coupling between $M_-$ magnon and $P_i=1,2,3$ phonon modes, the magnon absorption resonance dominates over those associated to $P_i$=1,2,3-phonons; these phonon resonances gain oscillator strength when they hybridize efficiently with the $M_-$ magnon branch at high magnetic fields. It is tempting to speculate that the effective spin-phonon coupling is at the origin of the above listed observations. This coupling, evidenced here for the characteristic $M$ and $P_i$ excitations, may also be thought to affect the ground state of our antiferromagnetic system, leading to a deformation of the unit cell that breaks its inversion symmetry. With such an assumption, the magnon-excitation can couple to light within the electric-dipole approximation what would account for our observation of a relatively strong magnon resonance in FIR transmission spectra. The $P_i$ phonon excitations apparent in Raman scattering, are then presumed to gain oscillator strength via their coupling to the magnon mode. This is in overall agreement with the evolution of the intensities of the $P_i$ absorption resonances when tuning the strength of the magnon-phonon coupling with the applied magnetic field. In fact, the only phonons which are observed to effectively couple to the magnon excitation are those apparent in both Raman scattering and FIR transmission spectra. The above speculations call for their solid verification on the theoretical ground. Nevertheless, our experimental demonstration of the effective optical activity of hybrid magnon-phonon modes in FePS$_3$ antiferromagnet may already be of special importance for future studies of this material in magnon/phonon optical pumping experiments~\cite{HayashiPRL18}.

\section{Conclusions}

In conclusion, we have uncovered the efficient interaction between the magnon and selective phonon modes in FePS$_3$, an archetype of van der Waals, quasi two-dimensional antiferromagnet. This interaction is revealed with magneto-spectroscopy methods which are uniquely operational in the apparent regime of coupling between the magnon-excitation and the lower-in-energy optical-like phonon modes. The strength of the magnon-phonon coupling is estimated with the clear observation of hybrid magnon-phonon modes, the magnon polarons, when the magnon-gap is shrinked with the applied magnetic field, to intersect the otherwise bare phonon modes. The hybrid magnon-phonon modes in FePS$_3$ are efficiently traced with Raman scattering experiments but they also directly couple to light, raising the pronounced resonances in FIR transmission spectra. This can be expected to trigger further exploration of FePS$_3$, by probing the magnetization dynamics in this antiferromagnet with THz pulsed excitation, including the offered possibility to tune the strength of magnon-phonon with an applied magnetic field. We also believe that magneto-spectroscopy techniques might be promising in studies of other magnetically ordered systems in which, in particular, the magnon excitations are suspected to couple to lower energy phonons. Micro-Raman scattering techniques continue to offer an interesting possibility to study magnetism in the ultimate limit of laterally small and strictly two-dimensional systems.

\vspace{5mm}

\begin{acknowledgments}
We acknowledge the discussions with Charles Simon, Andrew R. Wildes, Andr\'es Sa\'ul, Krishnendu Sengupta, and Abhishek Dey. The work has been supported by: EU Graphene Flagship project, CNRS via IRP ``2D materials'' and the ATOMOPTO project (TEAM programme of the Foundation for Polish Science, co-financed by the EU within the ERDFund). S.M. is grateful to DST-INSPIRE for the fellowship. S.D. acknowledges the financial support from DST-SERB grant No. ECR/2017/002037 and UGC-DAE CSR grant No. CRS-M-279. S.D. also acknowledges support from the Technical Research Centre (TRC), IACS, Kolkata. M.V acknowledges the support by the Operational Program Research, Development and Education financed by European Structural and Investment Funds and the Czech Ministry of Education, Youth and Sports (Project MATFUN -- CZ.02.1.01/0.0/0.0/15\_003/0000487). The Polish participation in EMFL is supported by the DIR/WK/2018/07.
\end{acknowledgments}

\vspace{5mm}

{\bf Author contributions:} M.~Potemski, S.D., M.O., and C.F. conceived the project and designed the experiments. A.G., S.M. and D.V. prepared the samples and performed their initial characterization together with M.~Palit.  D.V., M.~Palit, J.W., A.D., P.K., A.G., and M.G. performed the experiments and analyzed the data. M.V. proposed and developed the phenomenological model, with contributions from S.G., M.O. and M.~Potemski. All authors discussed the results and actively commented on the manuscript written by M.~Potemski with the initial input from D.V., M.~Palit and S.D.

\bibliography{FePS3.bib}

\newpage

\begin{figure}[htp]
\includegraphics[page=1,trim = 17mm 17mm 17mm 17mm, width=1.0\textwidth,height=1.0\textheight]{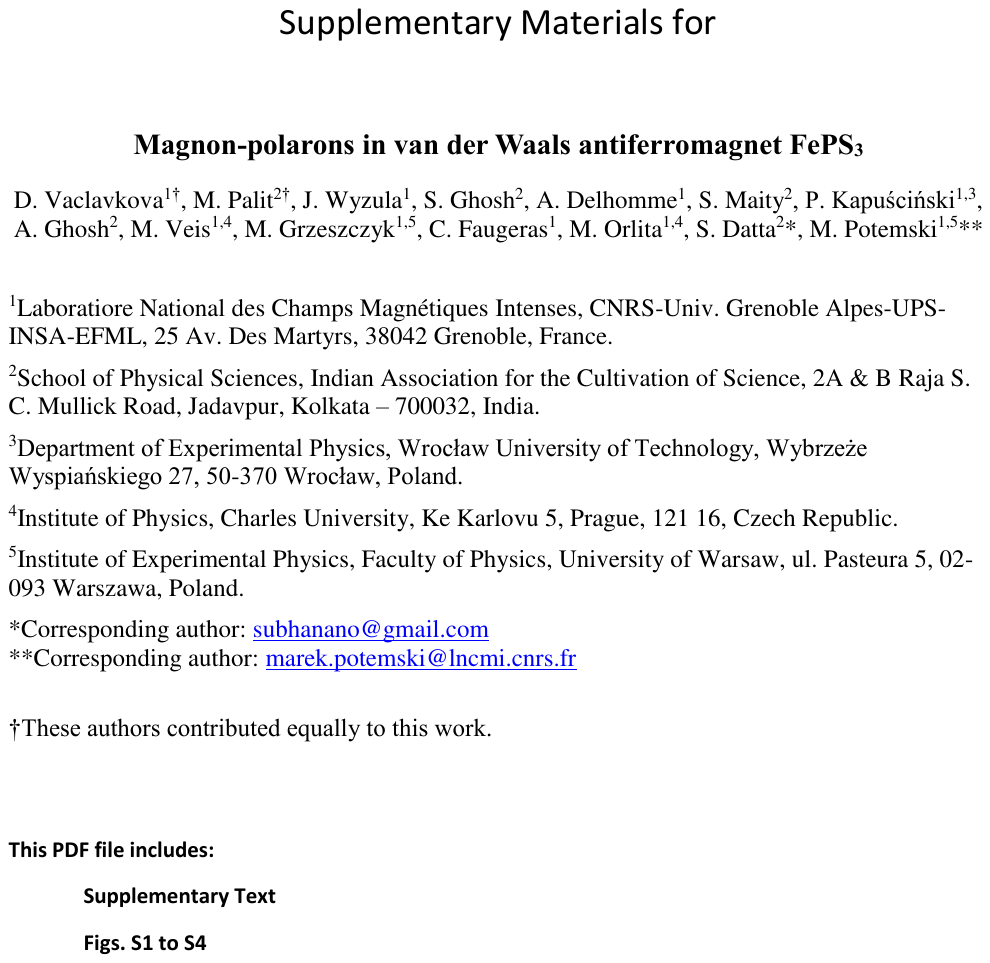}
\end{figure}

\newpage

\begin{figure}[htp]
  \includegraphics[page=2,trim = 17mm 17mm 17mm 17mm, width=1.0\textwidth,height=1.0\textheight]{Supplement.pdf}
\end{figure}

\newpage

\begin{figure}[htp]
  \includegraphics[page=3,trim = 17mm 17mm 17mm 17mm, width=1.0\textwidth,height=1.0\textheight]{Supplement.pdf}
\end{figure}

\newpage

\begin{figure}[htp]
  \includegraphics[page=4,trim = 17mm 17mm 17mm 17mm, width=1.0\textwidth,height=1.0\textheight]{Supplement.pdf}
\end{figure}

\newpage

\begin{figure}[htp]
  \includegraphics[page=5,trim = 17mm 17mm 17mm 17mm, width=1.0\textwidth,height=1.0\textheight]{Supplement.pdf}
\end{figure}

\end{document}